\documentclass{aastex631}
\usepackage[utf8]{inputenc}
\usepackage[T1]{fontenc}
\usepackage{listings}
\usepackage[deluxetable,revtex,bfvec]{math-shortcuts}
\hypersetup{colorlinks, linkcolor=blue, citecolor=blue, urlcolor=blue}
\begin{document}

\title{An unusual pulse shape change event in PSR J1713+0747 observed with the Green Bank Telescope and CHIME}
\author[0000-0003-1082-2342]{Ross J. Jennings}
\altaffiliation{NANOGrav Physics Frontiers Center Postdoctoral Fellow}
\affiliation{Department of Physics and Astronomy, West Virginia University, P.O. Box 6315, Morgantown, WV 26506, USA}
\affiliation{Center for Gravitational Waves and Cosmology, West Virginia University, Chestnut Ridge Research Building, Morgantown, WV 26505, USA}
\author[0000-0002-4049-1882]{James M. Cordes}
\author[0000-0002-2878-1502]{Shami Chatterjee}
\affiliation{Department of Astronomy, Cornell University, Ithaca, NY 14853, USA}
\affiliation{Cornell Center for Astrophysics and Planetary Science and Department of Astronomy, Cornell University, Ithaca, NY 14853, USA}
\author[0000-0001-7697-7422]{Maura A. McLaughlin}
\affiliation{Department of Physics and Astronomy, West Virginia University, P.O. Box 6315, Morgantown, WV 26506, USA}
\affiliation{Center for Gravitational Waves and Cosmology, West Virginia University, Chestnut Ridge Research Building, Morgantown, WV 26505, USA}
\author[0000-0002-6664-965X]{Paul B. Demorest}
\affiliation{National Radio Astronomy Observatory, 1003 Lopezville Road, Socorro, NM 87801, USA}
\author{Zaven Arzoumanian}
\affiliation{X-Ray Astrophysics Laboratory, NASA Goddard Space Flight Center, Code 662, Greenbelt, MD 20771, USA}
\author[0000-0003-2745-753X]{Paul T. Baker}
\affiliation{Department of Physics and Astronomy, Widener University, One University Place, Chester, PA 19013, USA}
\author[0000-0003-4046-884X]{Harsha Blumer}
\affiliation{Department of Physics and Astronomy, West Virginia University, P.O. Box 6315, Morgantown, WV 26506, USA}
\affiliation{Center for Gravitational Waves and Cosmology, West Virginia University, Chestnut Ridge Research Building, Morgantown, WV 26505, USA}
\author[0000-0003-3053-6538]{Paul R. Brook}
\affiliation{Department of Physics and Astronomy, West Virginia University, P.O. Box 6315, Morgantown, WV 26506, USA}
\affiliation{Center for Gravitational Waves and Cosmology, West Virginia University, Chestnut Ridge Research Building, Morgantown, WV 26505, USA}
\author[0000-0001-7587-5483]{Tyler Cohen}
\affiliation{Deptartment of Physics, New Mexico Institute of Mining and Technology, 801 Leroy Place, Socorro, NM 87801, USA}
\author[0000-0002-2578-0360]{Fronefield Crawford}
\affiliation{Department of Physics and Astronomy, Franklin \& Marshall College, P.O. Box 3003, Lancaster, PA 17604, USA}
\author[0000-0002-6039-692X]{H. Thankful Cromartie}
\altaffiliation{NASA Hubble Fellowship: Einstein Postdoctoral Fellow}
\affiliation{Cornell Center for Astrophysics and Planetary Science and Department of Astronomy, Cornell University, Ithaca, NY 14853, USA}
\author[0000-0002-2185-1790]{Megan E. DeCesar}
\affiliation{George Mason University, resident at the Naval Research Laboratory, Washington, DC 20375, USA}
\author[0000-0001-8885-6388]{Timothy Dolch}
\affiliation{Department of Physics, Hillsdale College, 33 E. College Street, Hillsdale, MI 49242, USA}
\affiliation{Eureka Scientific, Inc., 2452 Delmer Street, Suite 100, Oakland, CA 94602-3017, USA}
\author[0000-0001-7828-7708]{Elizabeth C. Ferrara}
\affiliation{Department of Astronomy, University of Maryland, College Park, MD 20742, USA}
\affiliation{Center for Research and Exploration in Space Science and Technology, NASA/GSFC, Greenbelt, MD 20771, USA}
\affiliation{Goddard Space Flight Center, Greenbelt, MD 20771, USA}
\author[0000-0001-8384-5049]{Emmanuel Fonseca}
\affiliation{Department of Physics and Astronomy, West Virginia University, P.O. Box 6315, Morgantown, WV 26506, USA}
\affiliation{Center for Gravitational Waves and Cosmology, West Virginia University, Chestnut Ridge Research Building, Morgantown, WV 26505, USA}
\author[0000-0003-1884-348X]{Deborah C. Good}
\affiliation{Department of Physics, University of Connecticut, 196 Auditorium Road, U-3046, Storrs, CT 06269-3046, USA}
\affiliation{Center for Computational Astrophysics, Flatiron Institute, 162 5th Avenue, New York, NY 10010, USA}
\author[0000-0003-2742-3321]{Jeffrey S. Hazboun}
\affiliation{University of Washington Bothell, 18115 Campus Way NE, Bothell, WA 98011, USA}
\author[0000-0001-6607-3710]{Megan L. Jones}
\affiliation{Center for Gravitation, Cosmology and Astrophysics, Department of Physics, University of Wisconsin-Milwaukee, P.O. Box 413, Milwaukee, WI 53201, USA}
\author[0000-0001-6295-2881]{David L. Kaplan}
\affiliation{Center for Gravitation, Cosmology and Astrophysics, Department of Physics, University of Wisconsin-Milwaukee, P.O. Box 413, Milwaukee, WI 53201, USA}
\author[0000-0003-0721-651X]{Michael T. Lam}
\affiliation{School of Physics and Astronomy, Rochester Institute of Technology, Rochester, NY 14623, USA}
\affiliation{Laboratory for Multiwavelength Astrophysics, Rochester Institute of Technology, Rochester, NY 14623, USA}
\author{T. Joseph W. Lazio}
\affiliation{Jet Propulsion Laboratory, California Institute of Technology, 4800 Oak Grove Drive, Pasadena, CA 91109, USA}
\author[0000-0003-1301-966X]{Duncan R. Lorimer}
\affiliation{Department of Physics and Astronomy, West Virginia University, P.O. Box 6315, Morgantown, WV 26506, USA}
\affiliation{Center for Gravitational Waves and Cosmology, West Virginia University, Chestnut Ridge Research Building, Morgantown, WV 26505, USA}
\author[0000-0001-5373-5914]{Jing Luo}
\altaffiliation{Deceased}
\affiliation{Department of Astronomy \& Astrophysics, University of Toronto, 50 Saint George Street, Toronto, ON M5S 3H4, Canada}
\author[0000-0001-5229-7430]{Ryan S. Lynch}
\affiliation{Green Bank Observatory, P.O. Box 2, Green Bank, WV 24944, USA}
\author[0000-0002-2885-8485]{James W. McKee}
\affiliation{Canadian Institute for Theoretical Astrophysics, University of Toronto, 60 Saint George Street, Toronto, ON M5S 3H8, Canada}
\affiliation{E.A. Milne Centre for Astrophysics, University of Hull, Cottingham Road, Kingston-upon-Hull, HU6 7RX, UK}
\affiliation{Centre of Excellence for Data Science, Artificial Intelligence and Modelling (DAIM), University of Hull, Cottingham Road, Kingston-upon-Hull, HU6 7RX, UK}
\author[0000-0003-2285-0404]{Dustin R. Madison}
\affiliation{Department of Physics, University of the Pacific, 3601 Pacific Avenue, Stockton, CA 95211, USA}
\author[0000-0001-8845-1225]{Bradley W. Meyers}
\affiliation{Department of Physics and Astronomy, University of British Columbia, 6224 Agricultural Road, Vancouver, BC V6T 1Z1, Canada}
\author[0000-0002-4307-1322]{Chiara M. F. Mingarelli}
\affiliation{Center for Computational Astrophysics, Flatiron Institute, 162 5th Avenue, New York, NY 10010, USA}
\affiliation{Department of Physics, University of Connecticut, 196 Auditorium Road, U-3046, Storrs, CT 06269-3046, USA}
\author[0000-0002-6709-2566]{David J. Nice}
\affiliation{Department of Physics, Lafayette College, Easton, PA 18042, USA}
\author[0000-0001-5465-2889]{Timothy T. Pennucci}
\affiliation{Institute of Physics, Eötvös Loránd University, Pázmány P.s. 1/A, 1117 Budapest, Hungary}
\author[0000-0002-8509-5947]{Benetge B. P. Perera}
\affiliation{Arecibo Observatory, University of Central Florida, HC3 Box 53995, Arecibo, PR 00612, USA}
\author[0000-0002-8826-1285]{Nihan S. Pol}
\affiliation{Department of Physics and Astronomy, Vanderbilt University, 2301 Vanderbilt Place, Nashville, TN 37235, USA}
\author[0000-0001-5799-9714]{Scott M. Ransom}
\affiliation{National Radio Astronomy Observatory, 520 Edgemont Road, Charlottesville, VA 22903, USA}
\author[0000-0002-5297-5278]{Paul S. Ray}
\affiliation{U.S. Naval Research Laboratory, Washington, DC 20375, USA}
\author[0000-0002-7283-1124]{Brent J. Shapiro-Albert}
\affiliation{Department of Physics and Astronomy, West Virginia University, P.O. Box 6315, Morgantown, WV 26506, USA}
\affiliation{Center for Gravitational Waves and Cosmology, West Virginia University, Chestnut Ridge Research Building, Morgantown, WV 26505, USA}
\affiliation{Giant Army, LLC, 915A 17th Ave, Seattle, WA 98122, USA}
\author[0000-0002-7778-2990]{Xavier Siemens}
\affiliation{Department of Physics, Oregon State University, Corvallis, OR 97331, USA}
\affiliation{Center for Gravitation, Cosmology and Astrophysics, Department of Physics, University of Wisconsin-Milwaukee, P.O. Box 413, Milwaukee, WI 53201, USA}
\author[0000-0001-9784-8670]{Ingrid H. Stairs}
\affiliation{Department of Physics and Astronomy, University of British Columbia, 6224 Agricultural Road, Vancouver, BC V6T 1Z1, Canada}
\author[0000-0002-1797-3277]{Daniel R. Stinebring}
\affiliation{Department of Physics and Astronomy, Oberlin College, Oberlin, OH 44074, USA}
\author[0000-0002-1075-3837]{Joseph K. Swiggum}
\altaffiliation{NANOGrav Physics Frontiers Center Postdoctoral Fellow}
\affiliation{Department of Physics, Lafayette College, Easton, PA 18042, USA}
\author[0000-0001-7509-0117]{Chia Min Tan}
\affiliation{Department of Physics, McGill University, 3600 rue University, Montréal, QC H3A 2T8, Canada}
\affiliation{McGill Space Institute, McGill University, 3550 rue University, Montréal, QC H3A 2A7, Canada}
\author[0000-0003-0264-1453]{Stephen R. Taylor}
\affiliation{Department of Physics and Astronomy, Vanderbilt University, 2301 Vanderbilt Place, Nashville, TN 37235, USA}
\author[0000-0003-4700-9072]{Sarah J. Vigeland}
\affiliation{Center for Gravitation, Cosmology and Astrophysics, Department of Physics, University of Wisconsin-Milwaukee, P.O. Box 413, Milwaukee, WI 53201, USA}
\author[0000-0002-6020-9274]{Caitlin A. Witt}
\affiliation{Center for Interdisciplinary Exploration and Research in Astrophysics, Northwestern University, 1800 Sherman Ave, Evanston, IL 60201}
\affiliation{Adler Planetarium, 1300 S. DuSable Lake Shore Drive, Chicago, IL 60605}
\date{\today}

\begin{abstract}
The millisecond pulsar J1713+0747 underwent a sudden and significant pulse shape change between April 16 and 17, 2021 (MJDs 59320 and 59321). Subsequently, the pulse shape gradually recovered over the course of several months. We report the results of continued multi-frequency radio observations of the pulsar made using the Canadian Hydrogen Intensity Mapping Experiment (CHIME) and the 100-meter Green Bank Telescope (GBT) in a three-year period encompassing the shape change event, between February 2020 and February 2023. As of February 2023, the pulse shape had returned to a state similar to that seen before the event, but with measurable changes remaining. The amplitude of the shape change and the accompanying TOA residuals display a strong non-monotonic dependence on radio frequency, demonstrating that the event is neither a glitch (the effects of which should be independent of radio frequency, $\nu$) nor a change in dispersion measure (DM) alone (which would produce a delay proportional to $\nu^{-2}$). However, it does bear some resemblance to the two previous ``chromatic timing events'' observed in J1713+0747 \citep{5yr-stochastic-background,lcc+16}, as well as to a similar event observed in PSR J1643$-$1224 in 2015 \citep{slk+16}.
\end{abstract}

\section{Introduction}

PSR J1713+0747 is a bright millisecond pulsar (MSP) that serves as a sensitive component of pulsar timing arrays (PTAs), projects that aim to detect gravitational waves at nanohertz frequencies through their influence on the arrival times of pulses from a large number of MSPs over many years. Its brightness (approximately 8.3 mJy at L band; \citealp{sbm+22}) and the sharp features in its pulse profile (effective width \SI{0.54}{ms} at L-band; based on data from \citealp{12yr-nb-timing}) make possible particularly high-precision time-of-arrival (TOA) measurements. As a result, J1713+0747 is currently observed by all member collaborations of the International Pulsar Timing Array (IPTA; \citealp{ipta-dr2}), namely, the North American Nanohertz Observatory for Gravitational Waves (NANOGrav; \citealp{12yr-nb-timing}; \citealp{12yr-wb-timing}), the European Pulsar Timing Array (EPTA; \citealp{dcl+16,epta-6psr}), the Parkes Pulsar Timing Array (PPTA; \citealp{krh+20}; \citealp{rsc+21}), and the Indian Pulsar Timing Array (InPTA; \citealp{jab+18,nag+22}), as well as by participating scientists associated with the Five-hundred-meter Aperture Spherical Telescope (FAST) in China \citep{fast-project,fast-pta} and the MeerKAT telescope in South Africa \citep{meertime}. In 2013,~\citet{dlc+14} undertook a 24-hour global observing campaign, using nine IPTA telescopes to assess the precision timing capabilities of PSR J1713+0747. They found that TOA measurement precision was ultimately limited by pulse jitter to a level of \SI{27}{\micro s} for single pulses, which would imply a TOA error of \SI{6.3}{ns} for the entire 24-hour observation. In addition to its use by PTAs, the high timing precision achievable with J1713+0747 and its well-characterized relativistic orbit make it an important pulsar for tests of general relativity \citep[e.g.][]{zdw+19}.

In April 2021, a large pulse shape change, noticeable by eye in comparisons of profiles, was detected in J1713+0737 \citep{xhb+21}. This change was surprising because the mean pulse profiles of pulsars are generally observed to be constant over long periods of time \citep[e.g.,][]{cordes13}. They presented profiles from before and after the shape change, observed with the Effelsberg and Nançay radio telescopes at C band (\num{4.9}--\SI{5.1}{GHz}) and L band (\num{1228}--\SI{1740}{MHz}), respectively, as well as with the Kunming 40-meter telescope at S band (\num{2150}--\SI{2400}{MHz}) and with FAST at L band (\num{1050}--\SI{1450}{MHz}). The FAST profiles included polarization information, indicating that the linearly polarized emission from the pulsar also changed as a result of the event. Additionally, \citet{xhb+21} noted that the shape change appeared to be accompanied by a change in the pulsar's dispersion measure (DM) of approximately \SI{-4.3e-3}{pc\,cm^{-3}}, which they measured using the FAST L-band data. 
The existence and timing of the event were subsequently confirmed using other telescopes, including the Giant Metrewave Radio Telescope (GMRT; \citealp{ssj+21}), the Green Bank Telescope (GBT; \citealp{lam21}), and the Canadian Hydrogen Intensity Mapping Experiment (CHIME; \citealp{mc21}). Since CHIME observes the pulsar with daily cadence, \citet{mc21} were able to constrain the time of the event's onset to a 24-hour period, between MJD 59320 and MJD 59321 (April 16 and 17, 2021). In the months after the shape change, the frequency-dependent pulse shape and arrival times have been observed to recover toward typical values seen prior to 2021 \citep{jennings22-atel}.

Importantly, the 2021 shape change event cannot be explained as a change in DM alone. A change in DM would not produce the complex changes in pulse shape that are observed, and the frequency dependence of the TOA residuals deviates significantly from the $\nu^{-2}$ scaling expected from dispersion.

On the other hand, this new shape change event in J1713+0747 is not entirely without precedent. PTA observations of J1713+0747 have shown two unusual chromatic (i.e., radio-frequency dependent) timing events in the last 15 years, although both are smaller in amplitude than the most recent event by approximately an order of magnitude. The first of these events, which took place around MJD 54750 (October 11, 2008), was identified by \citet{5yr-stochastic-background}. The second event took place around MJD 57510 (May 2, 2016), and was identified by \citet{leg+18}. Both events show an abrupt onset, lasting no more than a few days, after which pulses appear to arrive earlier. This is followed by a gradual, approximately exponential recovery taking place over several months. Since the TOA measurements change more at lower frequencies, each event produces a change in apparent DM. For the first event, this is about $\SI{-6e-4}{pc\,cm^{-3}}$, while for the second event, it is about $\SI{-4.3e-4}{pc\,cm^{-3}}$. \citet{leg+18} attribute the events to lensing of the radio emission by some structure in the ionized ISM. Follow-up by \citet{lll+21}, also using the NANOGrav 12.5-year data, showed that the data are consistent with a model in which the lensing is produced by a region of lower electron density with the geometry of a folded sheet. Shape changes associated with the first two chromatic timing events are not apparent by eye, but \citet{grs+21} and \citet{lll+21} have found that they do occur at a low level.

Another form of transient pulse shape change was observed in the Crab pulsar, PSR B0531+21, in 1997, where it was attributed to refractive lensing of the pulsar's radio emission by a sharp feature in the density distribution of plasma at the edge of the Crab nebula \citep{bwv00,glj11}. Given this context, it is apparent that there are many possible explanations for the event considered here. Several of these possibilities will be considered in more detail in Section~\ref{sec:interpretation} below.

In what follows, we report on recent observations of J1713+0747 made by NANOGrav and the CHIME/Pulsar Collaboration using the GBT and CHIME, both before and after the shape change event, and discuss some possible astrophysical scenarios which may have given rise to the event. In Section~\ref{sec:methods}, we describe the cadence and frequency coverage of the observations, the receivers used to make them, and the analysis techniques used to reduce them. In Section~\ref{sec:results}, we analyze the main features present in the data. Then, in Section~\ref{sec:interpretation}, we consider several possible astrophysical interpretations for the observed phenomena. Finally, in Section~\ref{sec:conclusions}, we summarize our results and conclude.

\section{Methods}\label{sec:methods}

\subsection{Observation}

\begin{deluxetable}{l c c c c c c c}
\label{table:obs-summary}
\tablecaption{Summary of observations}
\tablehead{\colhead{Receiver} & \multicolumn{2}{c}{Frequency (MHz)} & \multicolumn{2}{c}{MJD} & \colhead{Number} & \colhead{Cadence (d$^{-1}$)} & \colhead{Median S/N}\\
& \colhead{low} & \colhead{high} & \colhead{start} & \colhead{end} & & }
\startdata
GBT 1500 MHz & $1100$ & $1900$ & 58334 & 59945 & 136 & 0.084\tablenotemark{a} & 1410 \\
GBT 820 MHz & $720$ & $920$ & 58637 & 59916 & 37 & 0.029 & 938 \\
CHIME 600 MHz & $400$ & $800$ & 58596 & 59976 & 1096 & 0.794 & 65.6 \\
\enddata
\tablenotetext{a}{The average cadence of GBT 1500 MHz observations was reduced to equal that of the 820 MHz observations after March 2021 due to the end of the NANOGrav high-cadence observing program.}
\end{deluxetable}

The observations considered here were made using the GBT and CHIME telescopes. The GBT observations were made using two receivers: the \SI{800}{MHz} band of the PF1 prime focus receiver (PF1/800), which has a bandwidth of \SI{200}{MHz} centered on \SI{820}{MHz}, and the L-band Gregorian focus receiver, which has \SI{800}{MHz} of bandwidth centered on \SI{1500}{MHz}. The observations were coherently dedispersed and folded in real time using the VEGAS backend \citep{rbb+11,pbb+15} in its pulsar mode. Pulse profiles with 2048 phase bins were produced in initial channels \SI{1.5625}{MHz} wide, with 512 such channels across the band at \SI{1500}{MHz} and 128 at \SI{820}{MHz}. GBT observations at both \num{820} and \SI{1500}{MHz} were made roughly monthly as part of the main NANOGrav observing program, and typically have integration times of about 30 minutes. Additional \SI{1500}{MHz} observations of J1713+0747 were formerly made as part of the NANOGrav high-cadence observing program; unfortunately, these higher-cadence observations ended in March 2021, approximately a month before the shape change event began, so the average cadence of \SI{1500}{MHz} observations is lower after the onset of the shape change than before it. For a short (approximately 1 minute) period before each GBT observation, an artificial \SI{25}{Hz} pulsed noise source was injected into the signal path and recorded. These noise diode observations were later used to calibrate the differential gain and phase between the two hands of polarization of the receiver (cf. \citealp{9yr-dataset}, section 3.1).

The CHIME observations were made using the telescope's 256-antenna receiver system, which has \SI{400}{MHz} of bandwidth centered on \SI{600}{MHz}~\citep{chime}. Since CHIME consists of four stationary cylindrical reflectors oriented in a north-south direction, it observes a stripe along the meridian \SI{120}{\degree} long and \SI{1.3}{\degree}--\SI{2.5}{\degree} wide at any given moment, and cannot be pointed. Instead, the CHIME/Pulsar backend uses digitally synthesized beams to observe as many as ten pulsars simultaneously as they transit the main beam~\citep{chime-pulsar}. This makes it possible to observe nearly every pulsar visible to CHIME once every day, but only for a limited time, which for J1713+0747 is typically between 13 and 15 minutes. In practice, J1713+0747 was observed every day as it transited, except for the relatively few days when the telescope was shut down, either for maintenance or because of extreme high temperatures (more than \SI{45}{\celsius}). This gives the CHIME observations much higher cadence than even the ``high-cadence'' GBT observations. Like the GBT observations, the CHIME observations were also coherently dedispersed and folded in real time, to produce pulse profiles with 1024 phase bins in each of 1024 initial channels. However, due to the very different nature of observing with CHIME, noise diode observations were not available, and the polarization calibration performed for the GBT data could not be carried out (cf. \citealp{chime-pulsar}, section 4.4). For this reason, we do not attempt to interpret the polarization properties of the CHIME data below, but consider only the total intensity data.

A summary of all observations used in this paper is given in Table~\ref{table:obs-summary}. Below, we consider primarily those observations made between February 21, 2020 and February 1, 2023 (MJDs 58900--59976); observations from before this period were used only in constructing templates. GBT made a total of 29 observations with the \SI{820}{MHz} receiver and 70 observations with the \SI{1500}{MHz} receiver during this period. Of these, 16 of the \SI{820}{MHz} observations and 21 of the \SI{1500}{MHz} observations took place after the shape change event. In the same period, CHIME made 884 observations   of the pulsar, with 541 taking place after the shape change event. Due to the recentness of the event, none of the observations used in this paper are part of NANOGrav's published 12.5-year data set or the 15-year data set currently being prepared\footnote{The NANOGrav 15-year data set includes observations taken between July 30, 2004 (MJD 53216) and August 11, 2020 (MJD 59072) by two sets of pulsar backends: the Astronomical Signal Processor (ASP) at Arecibo and Green Bank Astronomical Signal Processor (GASP) at the GBT \citep{demorest-thesis}, and the later Green Bank Ultimate Pulsar Processing Instrument (GUPPI) at the GBT \citep{drd+08} and its sibling instruments PUPPI at Arecibo and YUPPI at the Very Large Array. GUPPI was decommissioned in April 2020 and replaced by the VEGAS backend. Some VEGAS observations used in this paper were made simultaneously with GUPPI observations that are included in the 15-year data set, but the VEGAS observations are not included in the 15-year data set, nor are the GUPPI observations used here. PUPPI observations ceased after a cable failure at Arecibo in August 2020, and the Arecibo telescope ultimately collapsed on December 1, 2020, meaning that they were never resumed.}.  However, all of them will be included in a future NANOGrav data set. Additionally, the full set of data used in this paper, including the frequency-resolved profiles derived from each observation, is available in an accompanying Zenodo dataset\footnote{\url{https://doi.org/10.5281/zenodo.7236460}}.

\subsection{Post-processing and alignment}\label{sec:alignment}

To prepare them for analysis, the GBT data were processed using NANOGrav's \texttt{nanopipe} software~\citep{nanopipe}, which excised pre-determined bands containing significant radio-frequency interference (RFI) or receiver resonances, and performed a basic polarization calibration procedure. No such pre-determined bands for RFI excision were available for CHIME, nor were the CHIME observations accompanied by dedicated calibration scans, so the CHIME data were not polarization calibrated, and all RFI removed was identified manually at a later step. All the data were then dedispersed at a constant reference DM of \SI{15.9638}{pc\,cm^{-3}} and aligned using a pulsar ephemeris fit to the NANOGrav 15-year data~\citep{15yr-timing}. The exact parameter values used can be found in the Zenodo dataset accompanying this paper. Finally, they were post-processed by removing portions of the band edges in which the bandpass filter rolled off significantly, as well as channels that were identified as containing significant (remaining) RFI. The presence of RFI was determined by manually inspecting the average profile for each receiver as a function of frequency and pulse phase.

Average profiles in the sixty-day periods immediately before and after the event, as well as in the most recent available data, are shown in Figure~\ref{fig:before-after-profiles}. We compared these profiles with templates for each band, created by averaging the profiles from all available observations made before the event. Then, for each profile shown in Figure~\ref{fig:before-after-profiles}, we subtracted a best-fit scaled, aligned copy of the corresponding template. In the presence of a shape change, the correct alignment between the profile and the template is ambiguous, because the normal procedure for fitting TOAs breaks down, as discussed in more detail below. To illustrate this, in quantifying the degree of the shape change, we used two different methods of alignment.

In the ``fixed-phase'' method, the template is aligned by extrapolating the phase predicted by a timing model fit to data before the event. As a result, the reference point for comparison in this case is the best prediction based on data available before the change. In effect, this method assumes that the ``true'' TOA (based on the rotation of the pulsar) has not changed as a result of the event. For this assumption to be a reasonable one, changes in the rotational phase of the pulsar due to intrinsic spin noise have to be small enough to be neglected. This is much more likely to be the case for an MSP such as J1713+0747 than for a canonical pulsar. In the ``fit-phase'' method, on the other hand, the template was aligned by fitting for a TOA in the ordinary manner (i.e., using Fourier-domain matched filtering, as described by \citealt{taylor92}), and shifting the template accordingly. This necessarily results in a smaller shape difference -- in fact, the smallest possible shape difference for any choice of alignment -- but there is no good reason to believe that the corresponding TOA is connected to the rotation of the pulsar. However, TOAs calculated in this way can serve as a useful point of comparison (see Section~\ref{sec:toa-dm} below). 

To quantify the degree to which the shape changed in each band, a fractional shape difference was derived for each combination of profile and alignment method. The results are shown in Table~\ref{table:shape-change-size} below. The shape difference was derived by first aligning the template with the profile, using either the ``fixed-phase'' or ``fit-phase'' method, and then fitting for the scale factor such that the aligned and scaled template (i.e., the profile model) most closely matched the observed profile. The model was then subtracted from the profile to produce a profile residual. The shape difference was then calculated by dividing the root-mean-square (RMS) of the profile residual by the RMS of the profile model (calculated after normalizing the profile to unit amplitude).

The frequency dependence of the profile shape, both before and after the event, can be seen more clearly in Figure~\ref{fig:before-after-portraits}, which shows the frequency-dependent profiles across all three bands in the same sixty-day periods used in Figure~\ref{fig:before-after-profiles}.  Figure~\ref{fig:profile-timeline} includes frequency-averaged profiles for the full set of GBT and CHIME data observed between February 21, 2020 and February 1, 2023, illustrating the time dependence of the profile shape. Because the frequency dependence of the profile shape is greatest in the CHIME band, Figure~\ref{fig:chime-subbands} is also included, demonstrating the time dependence of the profile shape in each of four \SI{100}{MHz} sub-bands of the CHIME data. To quantify the degree to which the shape had changed, we produced profile residuals by subtracting a best-fit scaled copy of the template from each of the profiles in Figure~\ref{fig:profile-timeline}. The results are shown in Figure~\ref{fig:residual-timeline}. In producing the profile residuals shown in Figure~\ref{fig:residual-timeline}, we used the ``fixed-phase'' method; i.e., the phase of the template subtracted from each profile was based on the timing model rather than fit to the data. As a result, the changes seen in Figure~\ref{fig:residual-timeline} are relative to expectations based on data from before the shape change.

\subsection{Principal component analysis}

We then performed principal component analysis (PCA) on the profile residuals.  
The vectors $v_{ik}$ in this basis are called the principal components. They are eigenvectors of the sample covariance matrix, $C_{ij}$, which has the eigendecomposition 
\begin{equation}C_{ij} = \frac1{n}\sum_{k=1}^nx_{ik}\mkern2mu x_{jk} = \sum_{\ell=1}^r \lambda_\ell\mkern2mu v_{i\ell}\mkern2mu v_{j\ell}.
\end{equation}
Here $n$ is the number of observations, and the indices $i$ and $j$ range over the $m$ phase bins. There are $r$ principal components, where $r=\operatorname{min}(m, n)$ is the rank of the covariance matrix, $C_{ij}$. The principal components are orthonormal in the sense that
\begin{equation}\label{eqn:orthonormal-pcs}
\sum_{i=0}^m v_{i\ell}\mkern2mu v_{i\ell'} = \delta_{\ell\ell'}.
\end{equation}
The eigenvalue, $\lambda_\ell$, corresponding to a principal component represents the amount of variance it describes in the data. Principal components are conventionally sorted in order of decreasing eigenvalue, and typically only the first few, most significant components are of interest.

Observations can be decomposed into principal components as
\begin{equation}\label{eqn:sum-of-pcs}
x_{ik} = \sum_{\ell=0}^r z_{k\ell}\mkern2mu v_{i\ell}.
\end{equation}
The coefficients, $z_{k\ell}$, in this expansion will be referred to as the principal component amplitudes. They represent the degree to which a particular principal component contributes to a particular observation, and their scale is determined by the normalization of the principal components. They can be computed from the data as
\begin{equation}\label{eqn:pc-amplitude-computation}
z_{k\ell}=\sum_{i=0}^m v_{i\ell}\mkern2mu x_{ik},
\end{equation}
which follows from the orthonormality of the principal components (equation~\ref{eqn:orthonormal-pcs}).

We performed PCA separately on the frequency-averaged profile residuals in each band, using a singular value decomposition of the profile residuals. The most significant principal components in each band (two in the GBT \SI{1500}{MHz} band, and three in each of the GBT \SI{820}{MHz} and CHIME bands) are shown in Figure~\ref{fig:pca-models}, along with the correspoding amplitudes, $z_{ij}$. No other principal components were significant. Multiplying the amplitudes by the corresponding principal components and adding them up, as in equation~(\ref{eqn:sum-of-pcs}), produces a time-dependent model for the shape of the profile, which will be discussed further in Section~\ref{sec:results} below. Uncertainties on the principal component amplitudes, $z_{ij}$, were calculated by assuming the observed profile residuals, $x_{ik}$, were subject to additive white noise. The amplitude of the noise in each profile was determined empirically from its discrete Fourier transform (DFT), by taking the standard deviation of the highest-frequency half of the Fourier components. Since the signal of interest is low-frequency, these highest frequency components consist almost entirely of noise, and, because the (appropriately normalized) DFT is a unitary transformation, the amplitude of this noise is directly related to the amplitude of the additive white noise in the time domain. The resulting uncertainty estimate for $x_{ij}$ was then propagated through equation ~\ref{eqn:pc-amplitude-computation}, assuming $z_{ij}$ is held fixed, to calculate the uncertainty on $z_{ij}$. This is a reasonably good estimate of the uncertainty in estimating the shape of the observed profile. However, it does not take into account any intrinsic variability in the true shape, which, as seen in section \ref{sec:results} below, may be significant.


\subsection{TOA and DM estimation}\label{sec:toa-dm}

Because standard TOA and DM estimates rely on the assumption that pulse shapes are consistent over time, they are necessarily biased in the presence of a pulse shape change. However, because such TOA and DM estimates are routinely produced as part of pulsar timing analyses, it is common for events like this one to be analyzed and quantified initially in terms of their effects on apparent TOA and DM. To facilitate comparison with such previous analyses, we produced TOA and DM estimates in the ordinary way; i.e., using the ``fit-phase'' procedure described above. As in the analyses described previously, the templates used were derived only from data collected prior to the onset of the shape change. This ensures that the TOA and DM estimates thus obtained, although unrelated to the true rotational phase of the pulsar or line-of-sight electron density, capture changes relative to the earlier, steady state.

More sophisticated TOA estimation techniques, which seek to correct for the effect of the shape change and recover the underlying rotation phase of the pulsar, are possible. For example, as shown by \citet{demorest-thesis} and further explored by \citet{ovh+11} in the context of pulse jitter, measured principal component amplitudes can be exploited to partially remove TOA estimation bias. Similar corrections can also be performed using other profile representations, including shapelet bases \citep{lah15,lkd+17} and bases derived from single-pulse clustering algorithms \citep{kerr15}. In this context, a caveat applies, in that if there is a true change in TOA associated with the pulse shape change and varying in amplitude together with it, such methods will also remove the TOA change. We intend to explore the use of these more sophisticated estimation techniques in future work, but here we consider only the na\"ive TOA and DM estimates described above.

To produce these estimates, we used the software package \texttt{PulsePortraiture} \citep{pdr15,pennucci19}. We derived a frequency-dependent profile model in each band by aligning and averaging the data observed prior to the event, performing PCA on the resulting per-channel profiles, and fitting a B-spline model to the corresponding principal component amplitudes, using the method described by \citet{pdr15}. In the frequency band between 720 and 800 MHz, which is visible both to the GBT 820 MHz receiver and to CHIME, the two profile models are largely consistent, with a shape difference (in the sense of Table~\ref{table:shape-change-size}) of less than 1.1\%. Using these models, we performed two types of fits to each observation. First, the full frequency-dependent profile model was fit to the data, producing a DM estimate and a single TOA estimate, referenced to the center frequency of the data. We refer to the TOA and DM estimates produced this way as ``wideband'' TOAs and DMs. Then, we divided each band into \SI{12.5}{MHz} channels and fit profile models derived from the full frequency-dependent model to the data in each channel separately, producing estimated TOAs in each channel. We refer to the results as ``narrowband'' TOAs. TOA residuals were produced from the narrowband TOAs by subtracting the average TOA in each frequency channel prior to the event from all of the TOAs in that channel, to eliminate the effect of any frequency dependence present prior to the shape change. Additionally, wideband DM residuals were produced by subtracting the mean DM in each receiver band prior to the shape change from the wideband DM estimates in that band. The TOA and DM residuals in each receiver band are shown in  Figure~\ref{fig:combined-timeline-plot}, and the TOA residuals are shown broken down by frequency in Figure~\ref{fig:frequency-dependence}. As mentioned above, the resulting TOA and DM measurements are highly covariant with the pulse shape, and should be understood as demonstrating how the event would appear in a conventional pulsar timing analysis, and not as describing true accompanying changes in the pulsar's spin or the column density of electrons along the line of sight.

\section{Results}\label{sec:results}

\begin{deluxetable}{l c c c c c c c}
\label{table:shape-change-size}
\tablecaption{Degree of shape change in each band. Fractional shape differences are calculated using the root-mean-square method described in Section~\ref{sec:alignment}.}
\tablehead{\colhead{Receiver} & \colhead{Epoch} & \colhead{MJD} & \multicolumn{2}{c}{Shape difference}\\
& & & \colhead{Fixed phase} & \colhead{Fit phase} }
\startdata
GBT 1500 MHz & onset & 59334 & $56.1\pm0.4\%$ & $26.7\pm0.3\%$ \\
GBT 820 MHz & onset & 59333 & $61.1\pm0.5\%$ & $24.5\pm0.4\%$ \\
CHIME 600 MHz & onset & 59321–81 & $42.3\pm0.6\%$ & $37.1\pm0.6\%$ \\
GBT 1500 MHz & most recent & 59945 & $7.08\pm0.32\%$ & $2.60\pm0.32\%$ \\
GBT 820 MHz & most recent & 59916 & $8.64\pm0.26\%$ & $2.89\pm0.26\%$\\
GBT 600 MHz & most recent & 59916–76 & $11.3\pm1.0\%$ & $10.1\pm1.0\%$ \\
\enddata
\end{deluxetable}

\begin{figure}
\centering
\includegraphics[width=0.9\textwidth]{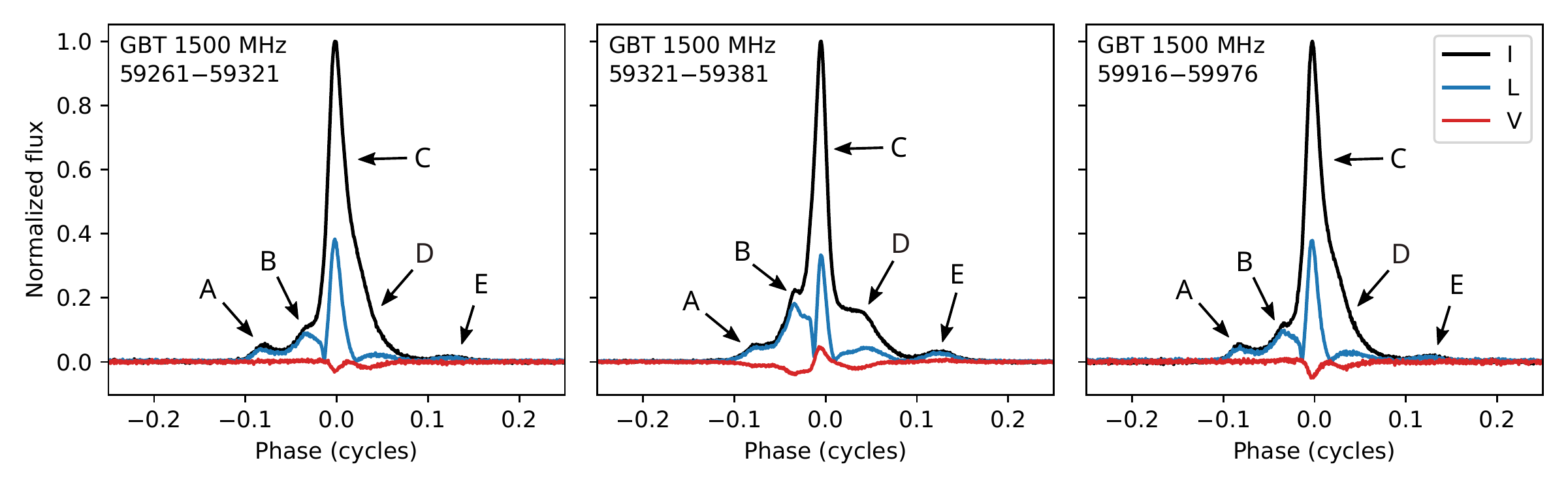}\vspace{-0.5em}
\includegraphics[width=0.9\textwidth]{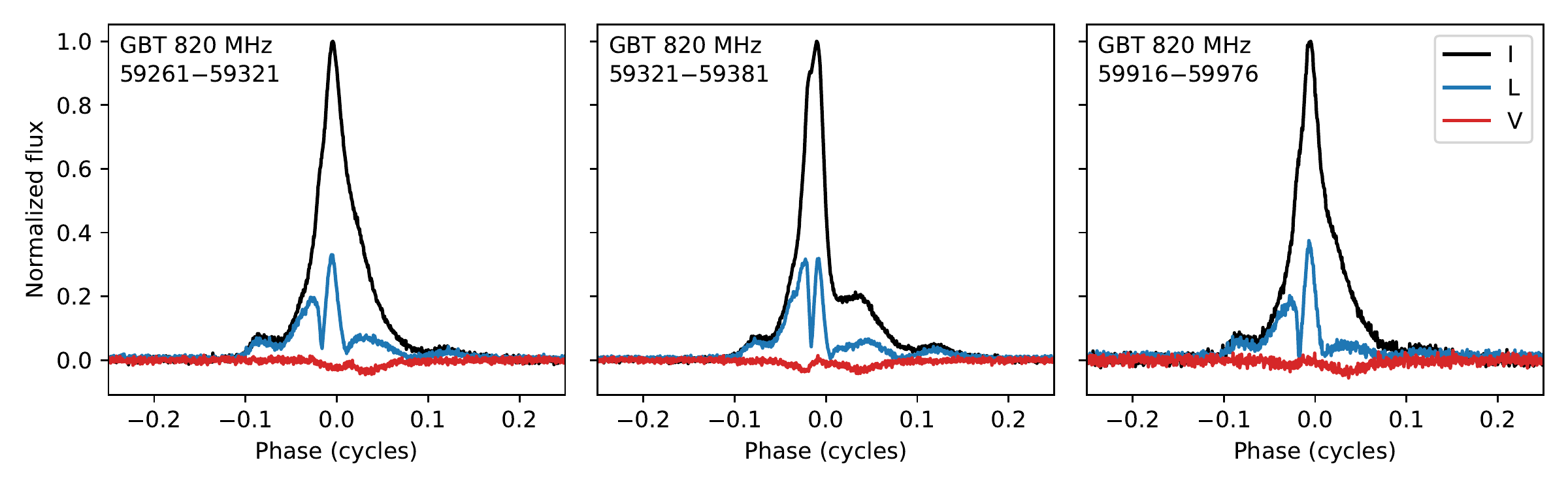}\vspace{-0.5em}
\includegraphics[width=0.9\textwidth]{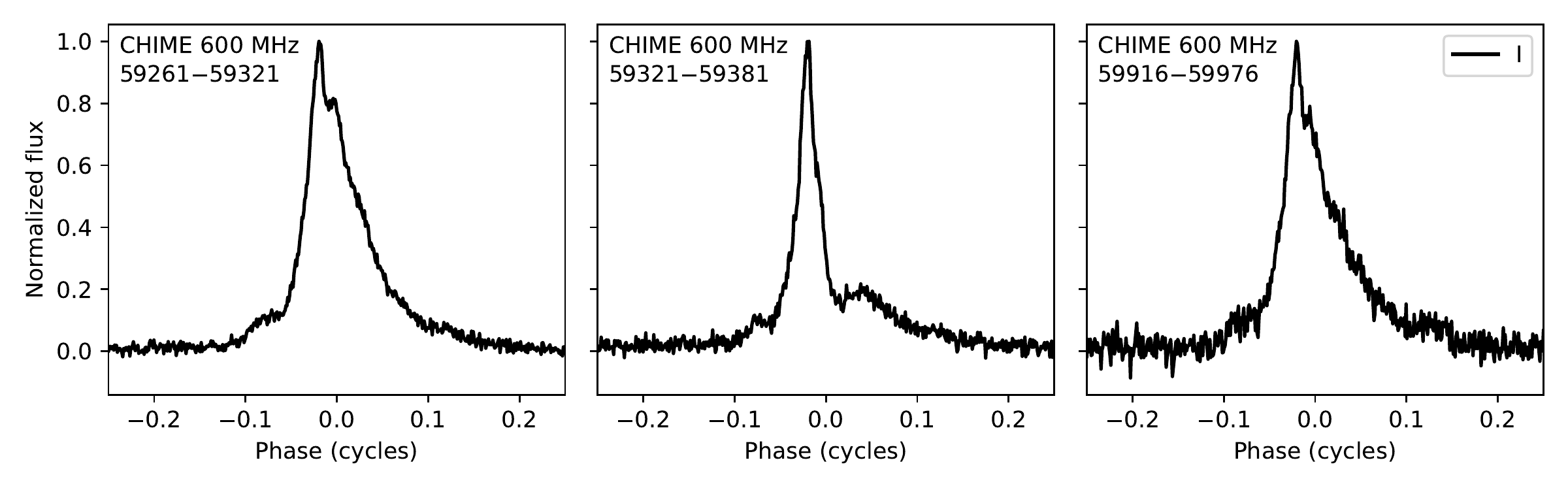}\vspace{-0.5em}
\caption{Average profiles before, during and after the event, as observed at GBT and CHIME. The left and center profiles are averaged over sixty-day periods immediately before (MJDs 59261--59321) and after the shape change (MJDs 59321--59381), while the right profile is averaged over a sixty-day period 595 days later (MJDs 59916--59976), after the pulse shape had mostly recovered. Each sixty-day period includes two GBT observations in each band and almost 60 CHIME observations. Profiles are averaged across the bands described in Table~\ref{table:obs-summary}. For the GBT data, the total linear polarization ($L$) and circular polarization ($V$) are shown in addition to the total intensity ($I$), while, for the CHIME data, which are not polarization calibrated, only the total intensity is shown.}
\label{fig:before-after-profiles}
\end{figure}


\begin{figure}
\centering
\includegraphics[width=\textwidth]{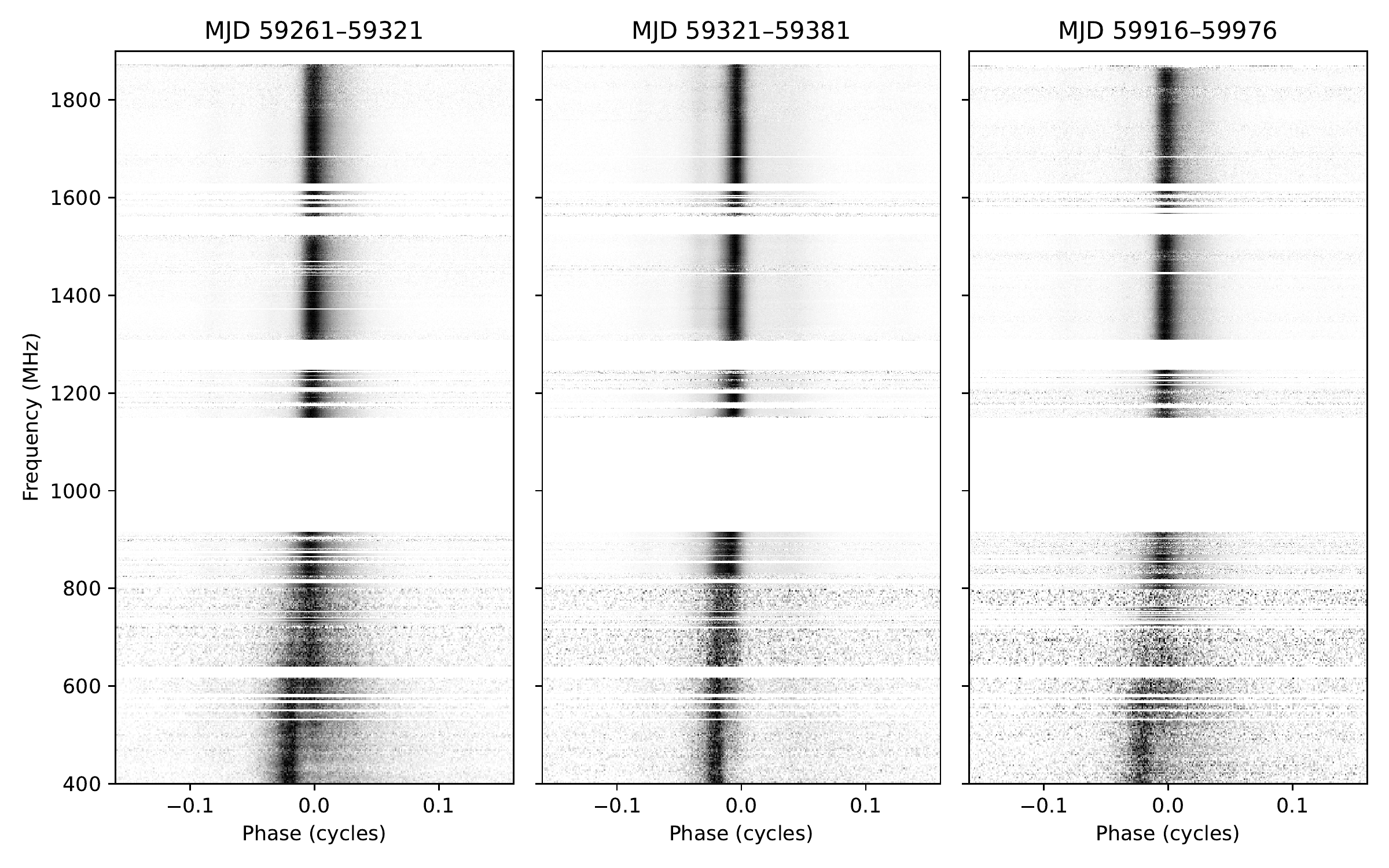}
\caption{Pulse profiles as a function of frequency and phase before, during, and after the shape change, averaged over the same sixty-day periods as in Figure~\ref{fig:before-after-profiles}. All three portraits are de-dispersed with the same dispersion measure, \SI{15.9638}{pc\,cm^{-3}}. Observations from all three receivers are included, with CHIME data plotted over GBT 820 MHz data where the bands overlap. A distinct leading component appearing only at the lowest frequencies is evident.}
\label{fig:before-after-portraits}
\end{figure}

\begin{figure}
\centering
\includegraphics[width=\textwidth]{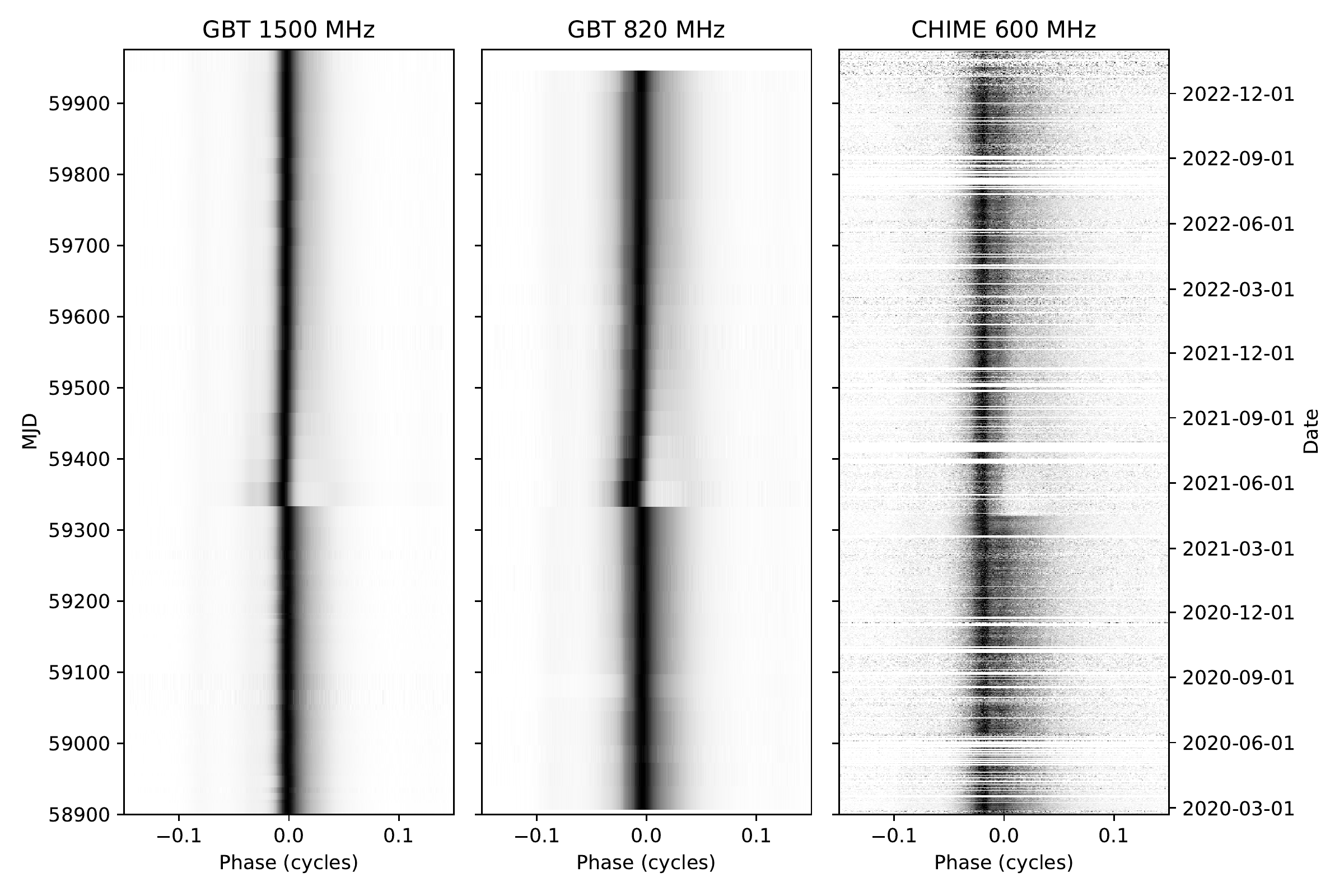}
\caption{Overview of the shape change event as seen by the GBT and CHIME. Profiles observed at GBT, using the VEGAS backend, are seen in the left two panels, and those observed at CHIME are shown in the right panel. All profiles are averaged across the bands described in Table~\ref{table:obs-summary}, and normalized to a constant peak intensity. Both GBT and CHIME observations are shown as horizontal bands whose lower edge corresponds to the observation date. For the CHIME observations, the band height is always one day, so gaps appear on days when CHIME did not observe the pulsar. For the much less frequent GBT observations, the bands are extended vertically until the date of the next observation, or, in the case of the last observation, for an additional 30 days. This means that the shape change appears to occur slightly earlier at CHIME than at GBT, when in fact, the lower observing cadence at GBT makes this impossible to determine.}
\label{fig:profile-timeline}
\end{figure}

\begin{figure}
\centering
\includegraphics[width=\textwidth]{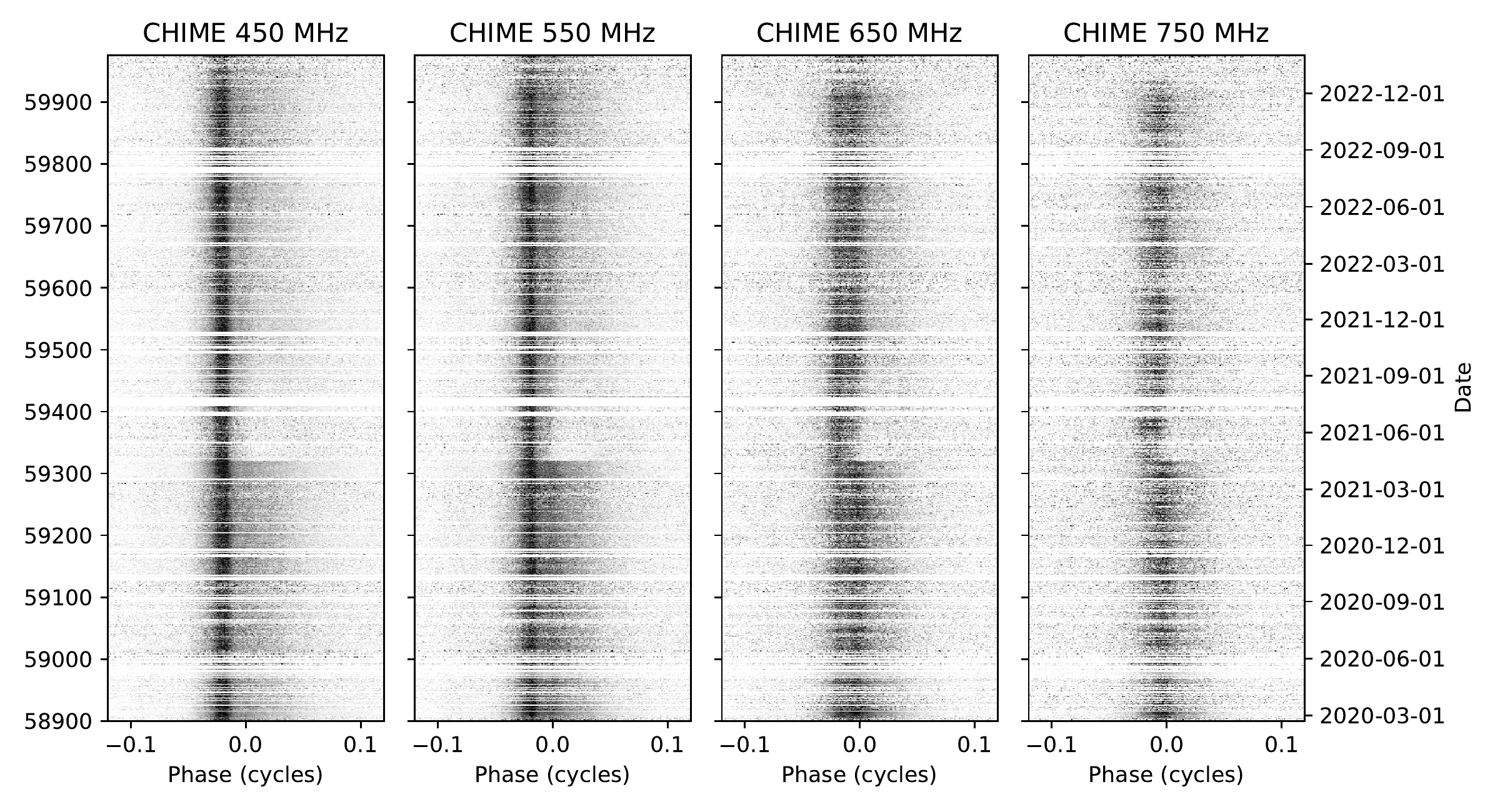}
\caption{The shape change event as seen in four 100 MHz sub-bands of the CHIME data. The low-frequency component seen in Figure~\ref{fig:before-after-portraits} is visible in the lower three sub-bands, and the high frequency component is visible in the upper three.}
\label{fig:chime-subbands}
\end{figure}

\begin{figure}
\centering
\includegraphics[width=\textwidth]{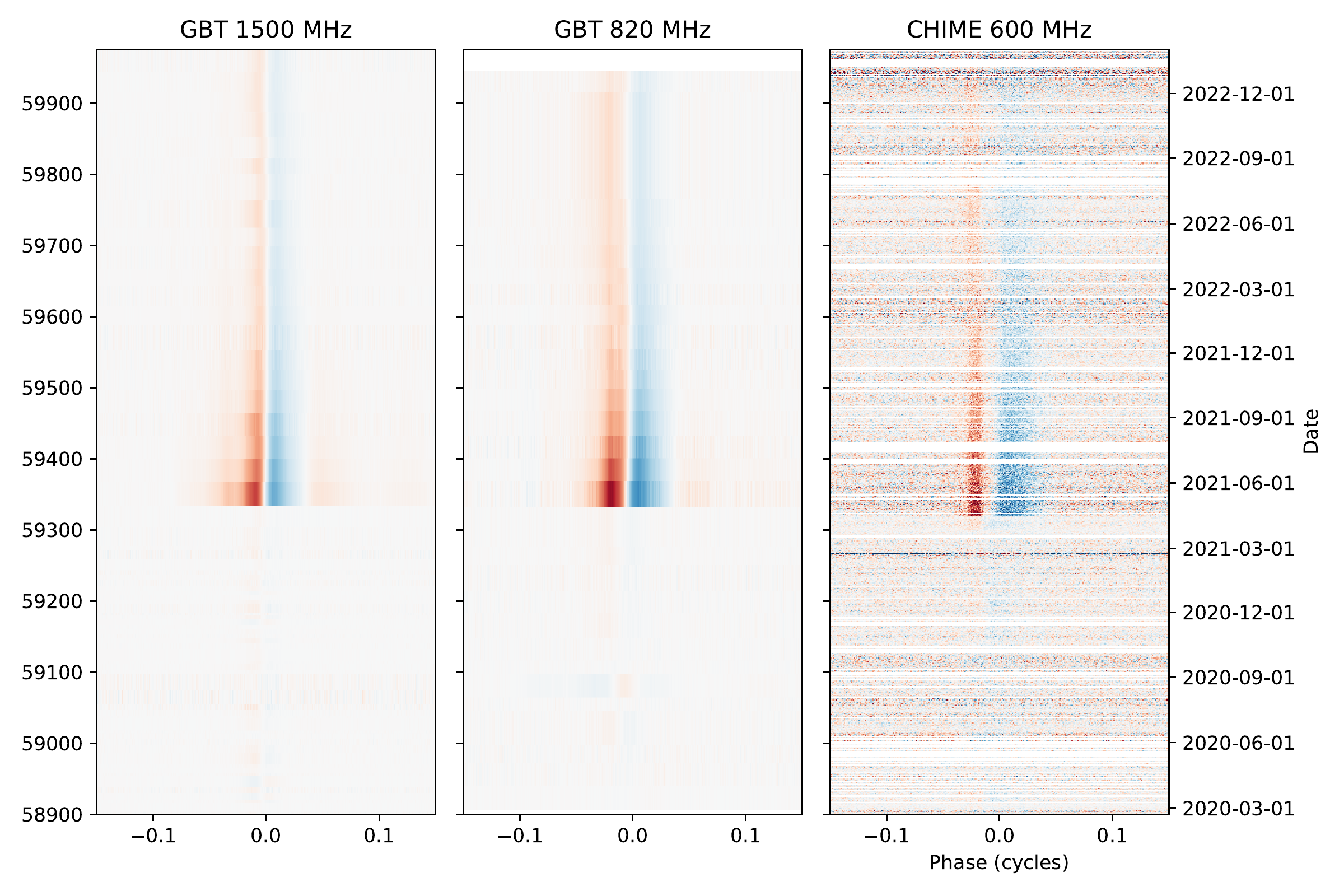}
\caption{Profile residuals around the shape change event, derived from the profiles in Figure~\ref{fig:profile-timeline} by subtracting a ``fixed-phase'' template (cf. Section~\ref{sec:alignment}). As in Figure~\ref{fig:profile-timeline}, GBT (VEGAS) observations are shown in the left two panels, and CHIME observations are shown in the right panel. Red represents positive residuals; i.e., regions where the observed intensity is greater than that predicted by the model, and blue represents negative residuals, regions where the observed intensity is less than the model prediction. Also, as in Figure~\ref{fig:profile-timeline}, the shape change appears to occur slightly earlier at CHIME than at GBT, but this is merely the result of the lower observing cadence at GBT.}
\label{fig:residual-timeline}
\end{figure}


\begin{figure}
\centering
\includegraphics[width=\textwidth]{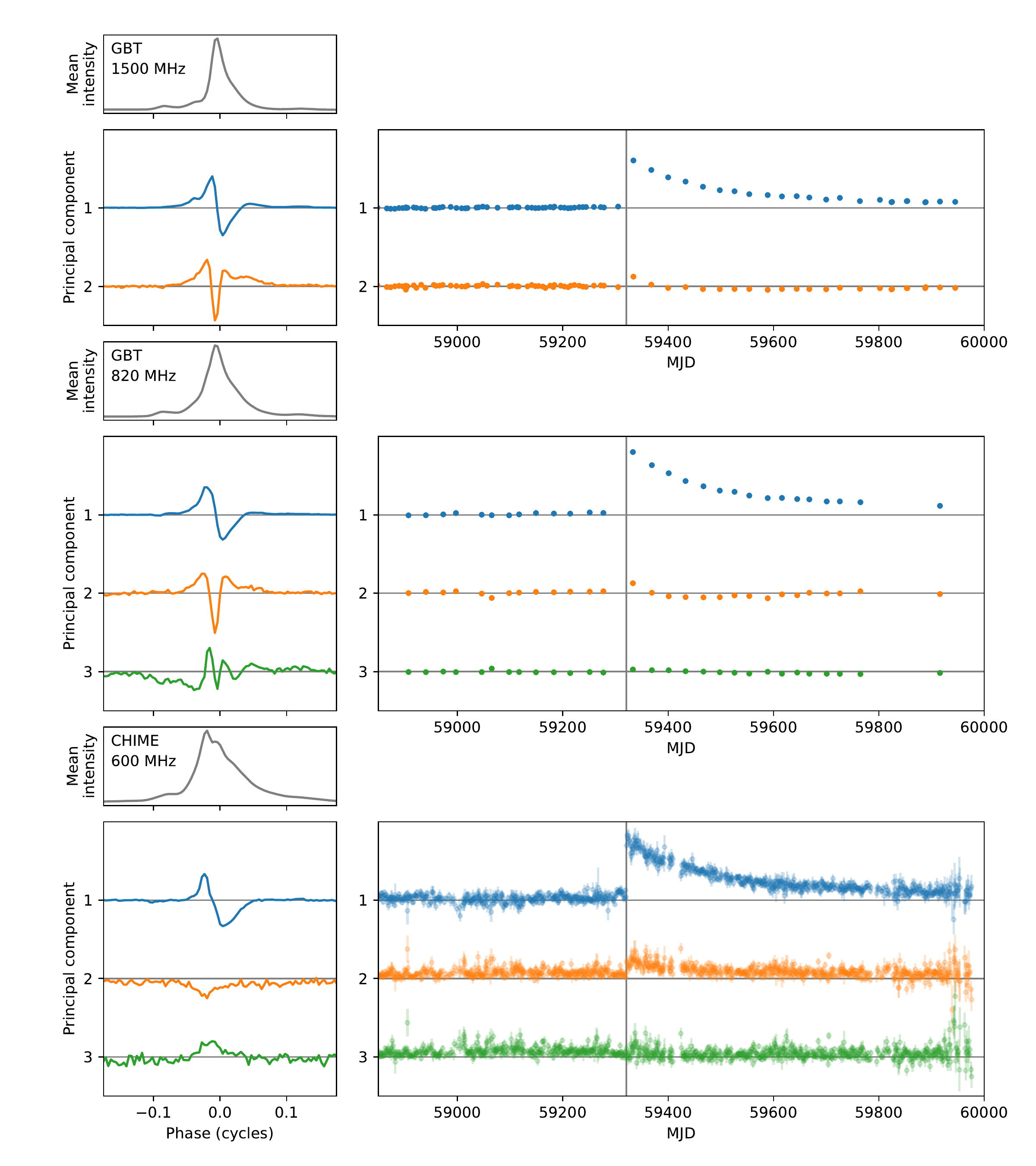}
\caption{Principal component analysis of the profile residuals in each band. For each of the three bands described in Table~\ref{table:obs-summary} (GBT 1500 MHz, GBT 820 MHz, CHIME 600 MHz), the three most significant principal components are shown at left, along with the average total intensity profile in that band. To the right, the corresponding principal component amplitudes are shown as a function of time. Adding the principal components, weighted according to their respective amplitudes, to the average profile gives a model of the profile shape at any particular observing epoch.}
\label{fig:pca-models}
\end{figure}

\begin{deluxetable}{l c c c c c c c}
\label{table:time-evolution}
\tablecaption{Time evolution of first principal component}
\tablehead{\colhead{Receiver} & \colhead{Model} & \colhead{$a$} & \colhead{$b$} & \colhead{$\tau$ (d)} & \colhead{$\alpha$} & \colhead{$\chi^2_r$}}
\startdata
& E & $1.05\pm0.06$ & \nodata & $271\pm12$ & \nodata & $512$ \\
GBT 1500 MHz & O & $1.14\pm0.04$ & $0.124\pm0.008$ & $156\pm8$ & \nodata & $111$ \\
& P & $1.36\pm0.05$ & \nodata & $190\pm40$ & $1.59\pm0.06$ & $95.7$ \\
\hline
& E & $1.42\pm0.08$ & \nodata & $253\pm17$ & \nodata & $357$ \\
GBT 820 MHz & O & $1.45\pm0.04$ & $0.271\pm0.015$ & $128\pm6$ & \nodata & $37.9$ \\
& P & $1.86\pm0.07$ & \nodata & $120\pm24$ & $1.17\pm0.11$ & $40.3$ \\
\hline
& E & $1.318\pm0.017$ & \nodata & $275\pm4$ & \nodata & $2.88$ \\
CHIME 600 MHz & O & $1.308\pm0.018$ & $0.206\pm0.009$ & $159\pm5$ & \nodata & $2.00$\\
& P & $1.580\pm0.026$ & \nodata & $204\pm23$ & $1.46\pm0.10$ & $1.95$\\
\enddata
\end{deluxetable}

\begin{figure}
\centering
\includegraphics[width=\textwidth]{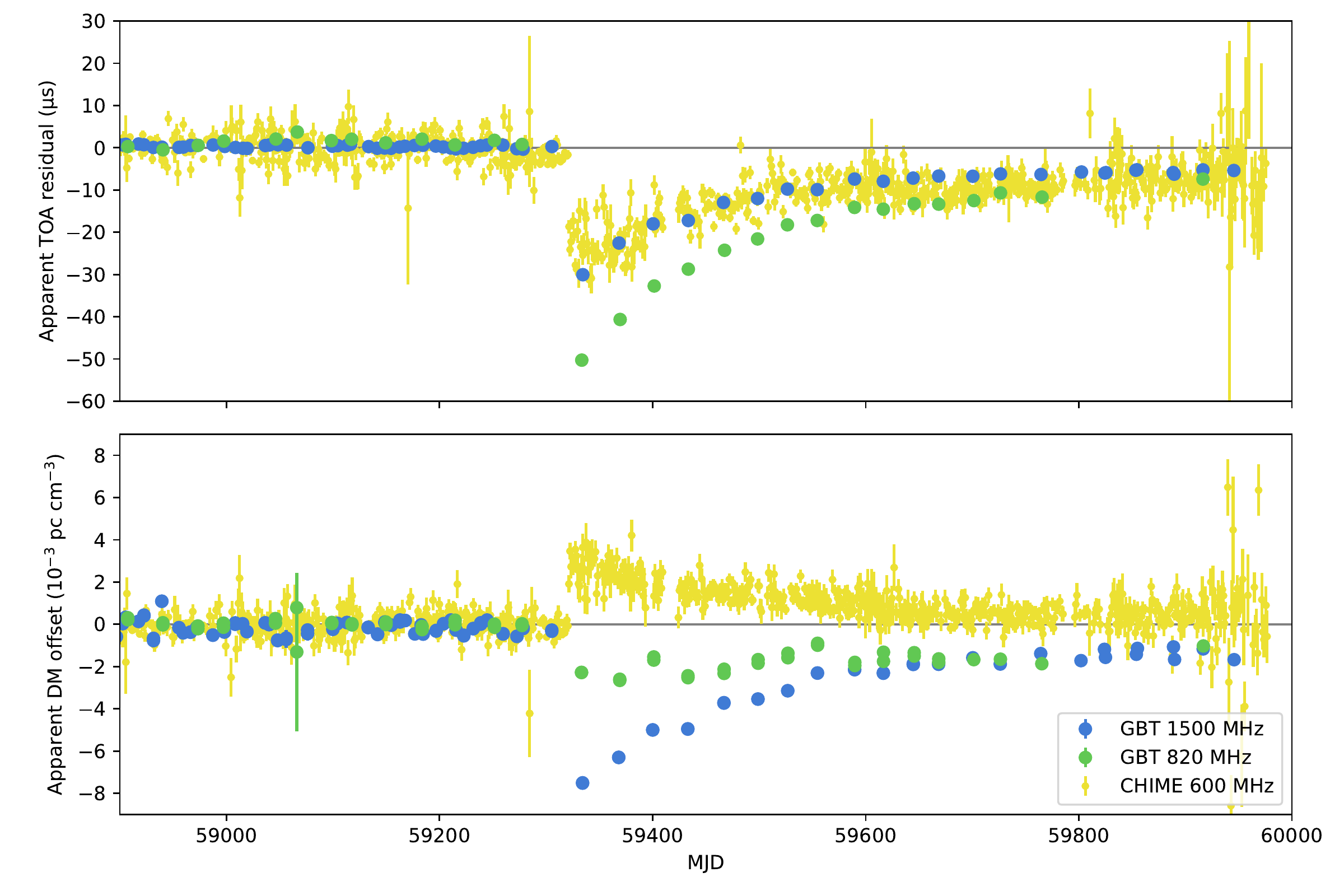}
\caption{Apparent band-averaged TOA (top) and DM (bottom) residuals before and after in the period around the shape change event, as measured using a profile model fit to data from before the event. Both the TOA and DM measurements are highly covariant with the pulse shape, and so these should be understood as demonstrating how the event would appear in a conventional pulsar timing analysis, and not as describing true accompanying changes in the pulsar's spin or the column density of electrons along the line of sight. The fact that the TOA residuals are a non-monotonic function of frequency, reaching their most negative value at around \SI{750}{MHz} (cf. Figure~\ref{fig:frequency-dependence}), can be seen in both of these plots: the amplitude of the TOA residuals is greater in the \SI{820}{MHz} band than at either \SI{600}{MHz} or \SI{1500}{MHz}; and the apparent DM residuals, which roughly represent the slope of the TOA residuals as a function of frequency, are positive at \SI{600}{MHz} and negative at \SI{1500}{MHz}.}
\label{fig:combined-timeline-plot}
\end{figure}

\begin{figure}
\centering
\includegraphics[width=\textwidth]{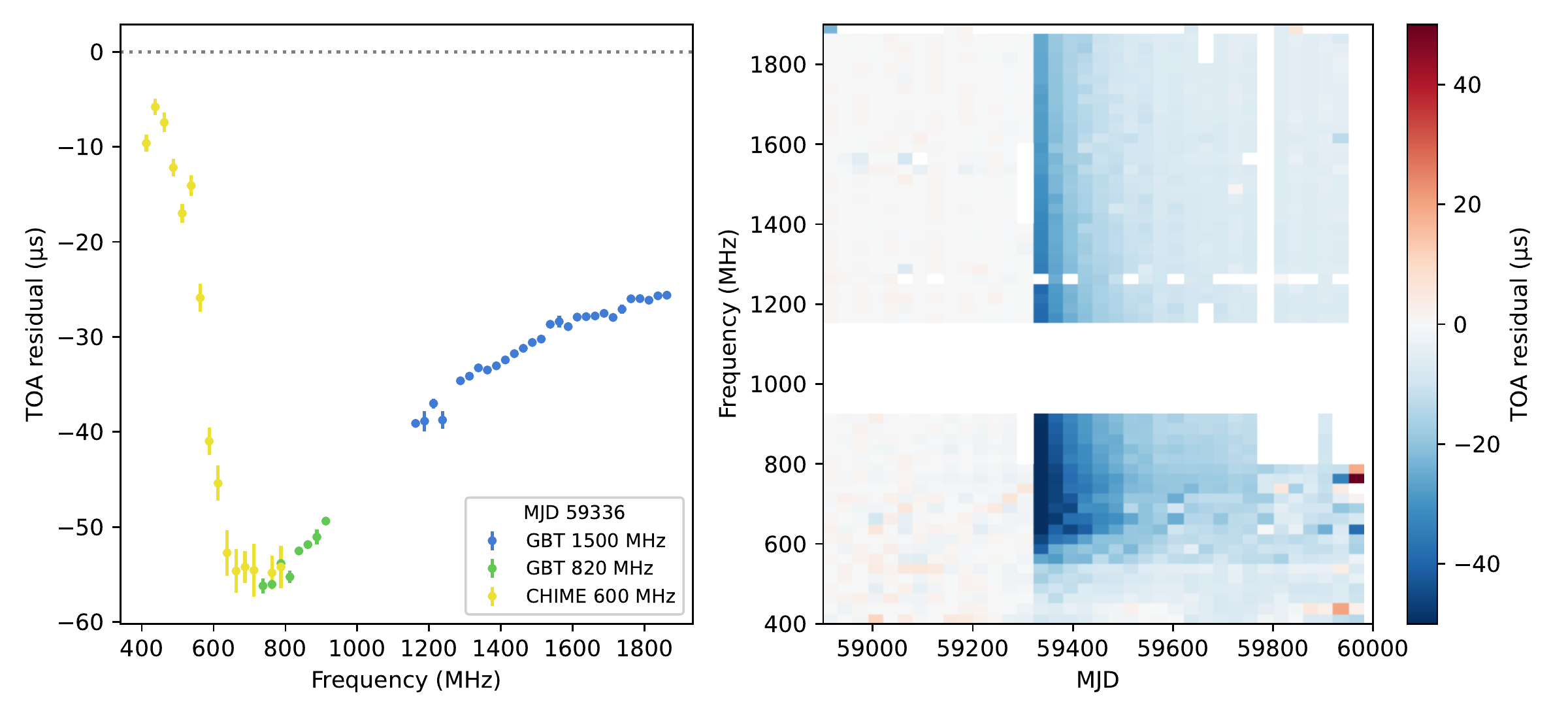}
\caption{Dependence of the apparent TOA residuals on frequency. The left panel shows the average TOA residual as a function of frequency, in each of two thirty day periods: immediately after the shape change (MJD 59321--59351, marked by dots), and at the end of the data presented here (MJD 59621--59651, marked by crosses). The right panel shows the TOA residual as a function of both time and frequency, averaged into \SI{25}{MHz}\,$\times$\,\SI{30}{day} bins. The frequency dependence is not consistent with the $\nu^{-2}$ form expected for dispersive behavior, but instead is non-monotonic, with the TOA residuals reaching their most negative value at around \SI{750}{MHz}.}
\label{fig:frequency-dependence}
\end{figure}

The average profiles seen in Figure~\ref{fig:before-after-profiles} illustrate the form of the shape change at different radio frequencies. As indicated in Table~\ref{table:shape-change-size}, all bands are affected to a similar degree: the shape difference is approximately 40 to 60 percent, compared to a fixed-phase template, or 20 to 40 percent compared to a template fit to the data in phase. Regardless of the alignment method used, this is much larger than the profile shape changes reported to accompany the 2016 event in this pulsar by \citet{lll+21}, which were approximately 1 to 4 percent of the peak profile amplitude.

Because the profile residuals were calculated relative to a fixed-phase template, any phase shift associated with the shape change event would have been absorbed into them, in the form of a shape change proportional to the derivative of the template shape (cf. \citealt{ovh+11}, section 4). Other kinds of shape changes may coincidentally align partially with the derivative of the template shape, but will in general have no relation to it. For the GBT 1500 MHz, GBT 820 MHz, and CHIME bands respectively, we find that 79.4\%, 85.9\%, and 36.6\% of the variance of the first principal component is in the direction of the derivative of the template shape, meaning that no more than roughly this fraction of the variance could be due to an associated phase shift.

Both before and after the onset of the event, the profile shows a noticeable dependence on frequency, as can be seen in Figure~\ref{fig:before-after-portraits}. Five main emission components can be identified: in addition to the bright central peak (C), there are two leading components (A, B) and two trailing components (D, E). One of these (D) is evident in the original profile only as a slight change in slope on the trailing edge of the main peak, together with a feature in the polarized emission, but appears much more distinct after the shape change. After the shape change, the main peak (C) is narrower, while the leading ``shoulder'' component (B) is brighter, and the trailing shoulder (D) is more pronounced. The second trailing component (E) also appears to be brighter than previously, when compared to the other components. Notably, the overall amplitude of emission from the pulsar is expected to vary between observing epochs as a result of scintillation, so it is much more difficult to measure absolute changes in the brightness of particular components than it is to measure relative changes in amplitude between the components. 

Figure~\ref{fig:profile-timeline} demonstrates the fact that the profile has recovered toward its original shape over the months that have elapsed since the event. At \SI{1.5}{GHz}, the leading shoulder component is brightest, and the trailing shoulder most distinct, immediately after the shape change, with both effects tending to diminish over time. At \SI{820}{MHz}, the leading shoulder appears to have briefly become approximately as bright as the main peak, and gradually declines in brightness over time, while, as at \SI{1.5}{GHz}, the trailing shoulder blends more and more with the main pulse. The recovery of the profile toward its original state is also clear in Figure~\ref{fig:residual-timeline}, which shows the profile residuals. The residuals can be seen to gradually decrease over time, while approximately maintaining their shape. As seen in Figure~\ref{fig:pca-models}, this recovery can be described primarily as a change in the amplitude of the first principal component, $z_1$, as a function of epoch. The form and timescale of the recovery can be quantified by fitting a parametric model to $z_1(t)$. We fit three such models to the data in each band: a decaying exponential (Model E):
\begin{equation}
    z_1(t)=a e^{-(t-t_0)/\tau},
\end{equation}
a decaying exponential with persistent offset (Model O):
\begin{equation}
    z_1(t) = ae^{-(t-t_0)/\tau} + b,
\end{equation}
and a power-law model (Model P):
\begin{equation}
    z_1(t) = a\sbrack{1+\frac{t-t_0}{\tau}}^{-\alpha}.
\end{equation}
The results can be seen in Table~\ref{table:time-evolution}. In each case, $t_0$ was taken to be MJD 59321.0 exactly; the true onset of the shape is uncertain, but differs from this by no more than about 12 hours. Data from before this point were excluded from the fit. Model E, the simplest, did not produce a satisfactory fit, but models O and P fit approximately equally well. Even for these better-fitting models, the resulting reduced chi-squared statistic was very large, reflecting the fact that there is additional short-time variability not accounted for by the nominal uncertainties. This is perhaps not surprising, since these uncertainties include only estimation error arising from white noise in the profile residuals, and not any intrinsic variability in pulse shape. Nevertheless, models O and P appear to describe the long-term evolution of the first principal component amplitude in each band relatively well. Note that model O predicts that some fraction of the shape change will remain indefinitely, whereas model P predicts that the profile shape will continue to recover, albeit more slowly. Unfortunately, since the data do not exhibit a clear preference between the two models, it is not possible to draw firm conclusions as to which of these predictions is more likely to hold.

The effect of the shape change on TOA estimates depends on frequency in a non-monotonic fashion, as demonstrated in  Figure~\ref{fig:combined-timeline-plot}. In the GBT \SI{820}{MHz} and \SI{1.5}{GHz} data, the effect of the shape change is consistently larger at lower frequencies. This frequency dependence manifests itself as an apparent change in DM peaking at approximately \SI{-7.5e-3}{pc\,cm^{-3}} in the \SI{1.5}{GHz} band and \SI{-2.5e-3}{pc\,cm^{-3}} in the \SI{820}{MHz} band, an order of magnitude larger than that seen in the previous two, smaller chromatic timing events, but in the same direction. This behaviour is consistent with what would be expected if the event were caused by the passage of an under-dense region of the ISM through the line of sight, which is the explanation preferred by \citet{leg+18} and \citet{lll+21} for the two previous chromatic timing events. However, the CHIME data indicate that, at the lowest frequencies, this frequency dependence reverses itself. This means that the apparent change in DM measured in the CHIME band has the opposite sense, peaking at approximately \SI{+2.5e-3}{pc\,cm^{-3}}, something which is difficult to explain in a model which relies on changes to the ISM along the line of sight to the pulsar.

Importantly, conventional TOA and DM measurements like those made here are always covariant with pulse shape changes, which makes it impossible to measure an absolute TOA without making assumptions about the pulse shape, and similarly impossible to measure an absolute DM without making assumptions about the frequency dependence of the pulse shape. Because we make the deliberately naïve choice to use a profile model based only on data prior to the shape change, the TOA and DM measurements given in this section and in Figures~\ref{fig:combined-timeline-plot} and~\ref{fig:frequency-dependence} should be understood as an empirical means of characterizing the shape change, rather than as a reflection of real changes to the underlying ``true'' TOA (defined with respect to the rotation of the pulsar) or DM (i.e., the column density of electrons along the line of sight).

\section{Interpretation}\label{sec:interpretation}

In many respects, the current event resembles a larger version of the two previous chromatic timing events. All three correspond to abrupt decreases in apparent DM, at least at frequencies above approximately \SI{800}{MHz}, and all three show at least some evidence for associated profile shape changes, which are only gradually frequency-dependent. The decay time of the new event is approximately 156 days at L-band (see Table~\ref{table:time-evolution}), which can be compared with the 62 and 25 day decay times derived for the previous two events by \citet{leg+18}. While the new event takes significantly longer to decay away, the decay time is of a similar order of magnitude. It may therefore be reasonable to assume that the new event has the same physical origin as those previous events. 

The shape change is too strongly correlated in frequency and time to be explained by diffractive interstellar scintillation (DISS). DISS can lead to apparent changes in pulse shape when observations of a pulsar with a frequency-dependent pulse shape are integrated across a wide band. This happens because scintillation makes the signal at certain frequencies appear brighter than at other frequencies, which biases the shape of the integrated pulse toward the shape at the brightened frequencies. However, this can be ruled out as the cause of the shape changes seen here. For J1713+0747, the bandwidth, $\Delta\nu_d$, and time, $\Delta t_d$, over which the scintillation pattern decorrelates, have been measured to be approximately \SI{22}{MHz} and \SI{48}{min}, respectively~\citep{lmj+16,tmc+21}. As described in Section~\ref{sec:results} above, the observed shape change is correlated over much larger bandwidths (hundreds of MHz) and times (months) than this, and its magnitude exceeds the degree of profile variation with frequency observed prior to the event. Moreover, the shape change can be also be seen in subbands narrower than the \SI{22}{MHz} scintillation bandwidth, in a form similar to what is seen in the integrated profiles.

An event possibly analogous to the profile shape change described here was observed in the MSP J1643$-$1224 beginning around MJD 57080 (February 27, 2015; \citealp{slk+16}). Like the recent event in J1713+0747, the 2015 event in J1643$-$1224 has a rise time of no more than a few days, and subsequently decays over the course of several months. \citet{slk+16} attribute this event to a magnetospheric disturbance of unspecified nature. It also shows a significant dependence on radio frequency, but one which is inverted compared to that seen in the first two events in J1713+0747, with more pronounced shape changes, and correspondingly larger timing effects, seen at higher frequencies. In observations with the 64-meter Parkes radio telescope (Murriyang), \citet{slk+16} found that the change in the shape of J1643$-$1224's profile was most prominent at \SI{10}{cm} (\SI{3}{GHz}), less noticeable at \SI{20}{cm} (\SI{1.5}{GHz}), and undetectable at \SI{50}{cm} (\SI{600}{MHz}). Notably, however, \citet{bkm+18} examined GBT observations of J1643$-$1224 taken during the same time period, and found that the effect was stronger at \SI{800}{MHz} than at \SI{1.5}{GHz}. This may point to a non-monotonic frequency dependence for the J1643$-$1224 event, as is seen here for the recent J1713+0747 event.

Several different hypotheses may be entertained as to the origin of the shape change event. One possibility is that the event might be due to lensing of the radio emission by a discrete structure in the ionized ISM along the line of sight. This is the generally accepted explanation for the 1997 event observed in the Crab pulsar \citep{bwv00,glj11}, and is the explanation proposed by \citet{leg+18} and \citet{lll+21} for the previous two events seen in PSR J1713+0747. Plasma lensing is expected to produce radio frequency-dependent changes in TOA measurements, and can also produce multiple images which are delayed in time by amounts comparable to the width of the pulse (\num{0.1}--\SI{1}{ms}), creating apparent profile shape changes~\citep{lcc+16,cwh+17}. However, the complex morphology of the shape change, which includes a sharpening of parts of the trailing edge, is not easily explained as a combination of a small number of lensed images. The non-monotonic frequency dependence of the TOA residuals may also point away from an ISM interpretation, but it is far from a perfect indicator. Simple changes in DM produce timing delays that are proportional to $\nu^{-2}$, and, while scattering and frequency-dependent DM effects can modify this dependence, they typically do so only modestly. However, ray tracing through a plasma lens can produce more complex frequency-dependent effects: refraction occurs due to the gradient of DM across the lens, and caustics can be produced. \citet{cwh+17} studied these phenomena in the context of fast radio bursts, finding that a variety of effects were possible, including cases where the frequnency dependence of arrival times differs markedly from the $\nu^{-2}$ scaling expected from dispersion alone. While plasma lensing does not seem to be the most likely explanation for the present event, it cannot be ruled out entirely without a more detailed analysis.

Some previously observed profile shape changes have been caused by geodetic precession, but this can be ruled out for J1713+0747. In binary neutron stars such as PSR B1913+16 and the double pulsar, PSR J0737$-$3039A/B, geodetic precession can cause changes in the viewing geometry, leading to pulse shape changes~\citep{wrt89,kramer98,bkk+08}. However, J1713+0747 has a white dwarf companion in a wide orbit with a 67.8-day period~\citep{sns+05}, which places it too far away for geodetic precession to be significant (the expected precession rate is a mere \SI{2.2}{mas/yr}). Additionally, for geodetic precession to produce detectable effects, a pulsar's spin direction must be significantly misaligned with its orbit, and the process of mass transfer that leads to the formation of MSP-white dwarf binaries like the J1713+0747 system tends to drive spin-orbit alignment. In any case, the rapid onset of the shape change seen here is inconsistent with such an origin.

Many other pulsars exhibit mode-changing, in which the profile abruptly changes back and forth between 2 or more shapes (e.g. \citealp{backer70}, \citealp{www+21}, \citealp{msb+22}).  This phenomenon has been shown to extend to millisecond pulsars \citep{mvmp18}. Gradual recovery of the profile toward a previous state, as is seen in this case, has not previously been observed in mode-changing pulsars. However, in some cases, there is a pronounced relationship between the profile state and the spin evolution of the pulsar \citep{klo+06,lhk+10,ssw+22}.  In the case of PSR~B1828$-$11, the fraction of time spent in each of the two profile modes varies roughly periodically with time and is directly related to the local value of the spin-down rate \citep{slk+19}.  Some of these “magnetospheric switching” pulsars can have very long-lived runs in a given profile state (e.g. PSR B2035+36, \citealp{ssw+22}).  It is conceivable that the profile change observed here in PSR J1713+0747 is seen to “decay” back to its pre-event state because of slow changes in the fraction of time spent in each of two extreme modes.  Detailed inspection of high-time-resolution data will be needed to determine whether or not this is the case.

In some cases, profile shape changes might be expected to accompany pulsar glitches. Glitches are a type of abrupt spin-up event, thought to be symptoms of the rapid transfer of angular momentum from a pulsar's interior to its crust through the unpinning of superfluid vortices~\citep{ai75,hm15}. Although they are most often seen in young pulsars, two glitches have been observed in MSPs~\citep{cb04,mjs+16}. Glitches are sometimes followed by a period of months in which the pulsar's period and period derivative recover toward their pre-glitch values. The rapid onset and subsequent recovery over seen in the shape change event in J1713+0747 are somewhat reminiscent of this phenomenon. However, glitches are not usually assoicated with profile shape changes, and in the few cases where profile changes associated with glitches have been observed, this is not what they look like. 
\citet{wje11} observed additional intermittent pulse components in the aftermath of a glitch in PSR J1119$-$6127, and \citet{ksj13} observed a relationship between glitching and emission state changes in PSR J0742$-$2822. More recently, \citet{pdh+18} managed to observe the Vela pulsar, PSR B0833$-$45, continuously over a period including the onset of a glitch, and saw changes in a few single pulses at the moment the glitch occurred. None of these scenarios is a particularly good analog for what is seen here in J1713+0747. The primary feature of a glitch is an achromatic step in the pulsar's spin frequency, which is not present in the current case. Furthermore, the shape changes which have previously been seen to accompany glitches are either transient or intermittent, rather than being sustained over several months with gradual decay, as seen here.

A more exotic possibility is that the shape change resulted from the injection of an asteroid into the pulsar's magnetosphere. \citet{bkb+14} argue that a new emission component seen in the canonical pulsar PSR B0736$-$40 in 2006 \citep{krj+11} was caused by such an asteroid incursion. Asteroidal material which entered the pulsar's magnetosphere, altering the flow of particles and thus the radio emission, could result in observable changes to a pulsar's radio emission \citep{cs08}. However, this is relatively unlikely for MSPs such as PSR J1713+0747. Since the light-cylinder radius $r_{\mr{LC}}=cP$ is 100 to 1000 times smaller for MSPs than for CPs, an asteroid is much more likely to be destroyed before entering the magnetosphere \citep{cs08}.

Overall, while the origin of the shape change event in J1713+0747 remains far from clear, it seems most likely that it was caused by some kind of change that took place within the pulsar's magnetosphere, as ISM-based explanations cannot easily account for its complex morphology. It is sufficiently different from a glitch that it must be considered a separate phenomenon. If it indeed has the same origin as the two previous events seen in J1713+0747 and the 2015 event in J1643$-$1224, it is by far the largest such event ever observed, and may reveal important aspects of the general nature of events of this kind.

\section{Conclusions}\label{sec:conclusions}

The recent shape change observed in PSR J1713+0747 most likely originated in the pulsar's magnetosphere. In several ways, it resembles a larger version of the two previously observed chromatic timing events \citep{demorest-thesis,leg+18,lll+21}, which have previously been attributed to lensing of the pulsar emission by underdense regions in the ISM. As in these past events, the DM, as measured between \num{820} and \SI{1500}{MHz}, has decreased abruptly, and appears to be recovering toward its original value on a timescale of several months. Although it is frequency-dependent, the event shows non-monotonic behavior inconsistent with a simple change in DM, and is accompanied by pulse shape changes that are nearly achromatic. There is some evidence that the same may be true of the two previous events.

ISM propagation effects in pulsar timing typically produce chromatic changes in TOAs. However, it is difficult to produce complex shape changes in this way. Lensing may produce multiple images of the pulse that interfere with each other to produce an altered profile shape, but systematic changes in the widths and relative amplitudes of pulse components, such as those seen in the event described here, would be much more likely given a magnetospheric origin.

The new event bears some resemblance to the profile shape change seen in PSR J1643$-$1224 in early 2015 \citep{slk+16}, which is thought to have had a magnetospheric origin. In that case, the frequency dependence was inverted, with the shape changes being stronger at higher frequencies. The non-monotonic frequency dependence of the event described here, however, means that an analogous origin for the J1643 event is very likely. An event of similar form may also have occurred in PSR J1640+2224 in mid-2012 (Hazboun et al., in prep.). Events like these may represent an entirely new phenomenon that will complicate millisecond pulsar timing, including pulsar timing array searches for gravitational waves. Fully accounting for them will likely require the use of time-varying templates referenced to the same fiducial pulse phase, as was done for PSR J0737$-$3039B, the young pulsar component of the double pulsar, whose pulse changes shape as a result of geodetic precession, in~\citet{ndk+20}. We intend to explore the use of time-varying templates in more detail in future work.

The trend toward increasing fractional bandwidths in pulsar observing has created a need for pulsar timing techniques which can account for profile shape variations across a wide band, without imposing an undue computational burden on later analysis. Compared to splitting a wide band into many narrow subbands and generating traditional narrowband TOAs within each subband, wideband techniques like those of \citet{pennucci19} can account for such shape variations in a natural way while reducing the amount of data that must be processed in subsequent analysis steps to a single TOA and DM (rather than a TOA in each subband) per epoch. The downside of this approach is that other kinds of frequency dependence, such as time-variable scattering or the kind of time-and-frequency dependent event seen here, can no longer be dealt with at the level of TOA residuals, since information about frequency dependence within each band, beyond the DM, is lost. There is a potential solution to this problem, however, which is to deal with these frequency-dependent effects directly at the level of the profiles, by adding appropriate terms to the wideband TOA log-likelihood \citep[cf.][]{pennucci19,12yr-wb-timing}. Indeed, such an approach may be the only way to deal with an event like the one described here within the a fully wideband framework.

Events such as the present one will present a challenge for low-frequency gravitational wave searches by PTAs. However, there are good reasons to believe this challenge is surmountable. Gravitational wave signals are expected to be dominated by very low-frequency components, appear independent of radio frequency, and show a characteristic quadrupolar pattern of spatial correlations between pulsars, quantified by the Hellings \& Downs curve~\citep{hd83,cs13,12yr-stochastic}. Because of its limited extent in time (in particular, data taken before the onset of the event is unaffected by it), its dependence on radio frequency, and its appearance in only a single pulsar, this event is not likely to be strongly covariant with gravitational-wave signals, and it should in principle be possible to model and subtract its effects without severely affecting the gravitational-wave sensitivity of a PTA data set.\\
\\
The NANOGrav Physics Frontiers Center receives support from the National Science Foundation (NSF) under award number 1430284. The Green Bank Observatory is a facility of the NSF operated under cooperative agreement by Associated Universities, Inc. We acknowledge that CHIME is located on the traditional, ancestral, and unceded territory of the Syilx/Okanagan people. We are grateful to the staff of the Dominion Radio Astrophysical Observatory, which is operated by the National Research Council of Canada.  CHIME is funded by a grant from the Canada Foundation for Innovation (CFI) 2012 Leading Edge Fund (Project 31170) and by contributions from the provinces of British Columbia, Qu\'{e}bec and Ontario. The CHIME/FRB Project, which enabled development in common with the CHIME/Pulsar instrument, is funded by a grant from the CFI 2015 Innovation Fund (Project 33213) and by contributions from the provinces of British Columbia and Qu\'{e}bec, and by the Dunlap Institute for Astronomy and Astrophysics at the University of Toronto. Additional support was provided by the Canadian Institute for Advanced Research (CIFAR), McGill University and the McGill Space Institute thanks to the Trottier Family Foundation, and the University of British Columbia. The CHIME/Pulsar instrument hardware was funded by NSERC RTI-1 grant EQPEQ 458893-2014. This research was enabled in part by support provided by WestGrid (www.westgrid.ca) and Compute Canada (www.computecanada.ca).

Support for H. T. Cromartie is provided by NASA through the NASA  Hubble  Fellowship  Program  grant  \#HST-HF2-51453.001 awarded by the Space Telescope Science Institute, which is operated by the Association of Universities for Research in Astronomy, Inc., for NASA, under contract NAS5-26555. T. Dolch is supported by an NSF Astronomy and Astrophysics Grant (AAG) award number 2009468. E. C. Ferrara is supported by NASA under award number 80GSFC21M0002. M. T. Lam is supported by an NSF Astronomy and Astrophysics Grant (AAG) award number 2009468. The Flatiron Institute is supported by the Simons Foundation. Part of this research was carried out at the Jet Propulsion Laboratory, California Institute of Technology, under a contract with the National Aeronautics and Space Administration. J. W. McKee is a CITA Postdoctoral Fellow: This work was supported by Ontario Research Fund—research Excellence Program (ORF-RE) and the Natural Sciences and Engineering Research Council of Canada (NSERC) [funding reference CRD 523638-18]. T. T. Pennucci acknowledges support from the MTA-ELTE Extragalactic Astrophysics Research Group, funded by the Hungarian Academy of Sciences (Magyar Tudományos Akadémia), which was used during the development of this research. S. M. Ransom is a CIFAR Fellow. Pulsar research at UBC is supported by an NSERC Discovery Grant and by CIFAR. Portions of this work performed at NRL were supported by ONR 6.1 basic research funding.\\
\\
Facilities: Green Bank Observatory, CHIME\\
\\
Software: DSPSR \citep{dspsr:2010ascl,dspsr:2011pasa}, PSRCHIVE \citep{psrchive:2004pasa,psrchive:2011ascl}, \texttt{psrtools} (\url{https://github.com/demorest/psrtools}), \texttt{nanopipe} \citep{nanopipe}, Astropy \citep{astropy:2013,astropy:2018,astropy:2022}, PulsePortraiture \citep{pulseportraiture}, PINT \citep{pint}

\bibliography{main}

\end{document}